\documentclass[notitlepage, 10pt]{article}

\usepackage[top=2cm, bottom=2cm, left=1.25cm, right=1.25cm]{geometry}
\usepackage{graphicx}

\usepackage{titlesec}
\usepackage{sectsty}
\usepackage{authblk}
\usepackage{siunitx}
\usepackage{amsmath}
\usepackage[super,comma,sort&compress]{natbib}
\titlespacing\section{0pt}{2pt plus 4pt minus 2pt}{0pt plus 2pt minus 2pt}
\sectionfont{\fontsize{12}{15}\selectfont}
\paragraphfont{\fontsize{9}{9}\selectfont}
\usepackage{braket}
\usepackage[font=small,labelfont=bf]{caption}
\usepackage{lineno}
\usepackage{nameref}
\usepackage{newfloat}
\usepackage{xcolor}
\usepackage{mhchem}
\DeclareFloatingEnvironment{metfig}

\usepackage{hyperref}
\hypersetup{
    colorlinks=true,
    linkcolor=black,
    citecolor=black,
    filecolor=black,
    urlcolor=black,
}

\title{Coherent control of electron spin qubits in silicon using a global field}
\date{}
\author[1]{E. Vahapoglu}
\author[1]{J. P. Slack-Smith}
\author[1]{R. C. C. Leon}
\author[1]{W. H. Lim}
\author[1]{F. E. Hudson}
\author[1]{T. Day}
\author[1]{J. D. Cifuentes}
\author[1]{T. Tanttu}
\author[1]{C. H. Yang}
\author[1]{A. Saraiva}
\author[2]{N. V. Abrosimov}
\author[3]{H.-J. Pohl}
\author[4]{M. L. W. Thewalt}
\author[1]{A. Laucht}
\author[1,*]{A. S. Dzurak}
\author[1,*]{J. J. Pla}
\affil[1]{School of Electrical Engineering and Telecommunications, UNSW Sydney, Sydney, NSW 2052, Australia.}
\affil[2]{Leibniz-Institut für Kristallzüchtung, 12489 Berlin, Germany.}
\affil[3]{VITCON Projectconsult GmbH, 07745 Jena, Germany.}
\affil[4]{Department of Physics, Simon Fraser University, Burnaby, British Columbia V5A 1S6, Canada.}
\affil[*]{These authors contributed equally to this work}
\begin{document}

    \maketitle
    
    \begin{abstract}
Silicon spin qubits promise to leverage the extraordinary progress in silicon nanoelectronic device fabrication over the past half century to deliver large-scale quantum processors. Despite the scalability advantage of using silicon technology, realising a quantum computer with the millions of qubits required to run some of the most demanding quantum algorithms poses several outstanding challenges, including how to control so many qubits simultaneously. Recently, compact 3D microwave dielectric resonators were proposed as a way to deliver the magnetic fields for spin qubit control across an entire quantum chip using only a single microwave source. Although spin resonance of individual electrons in the globally applied microwave field was demonstrated, the spins were controlled incoherently. Here we report coherent Rabi oscillations of single electron spin qubits in a planar SiMOS quantum dot device using a global magnetic field generated off-chip. The observation of coherent qubit control driven by a dielectric resonator establishes a credible pathway to achieving large-scale control in a spin-based quantum computer.
	\end{abstract}

	\twocolumn
	
Quantum computers promise to transform our ability to solve currently intractable problems, with implications for critical sectors such as finance, defence and pharmaceuticals. The benefits of quantum computation are already expected to be observed with ``noisy intermediate scale quantum'' (NISQ) devices \cite{Preskill2018}, which possess tens to hundreds of qubits. However, it is widely acknowledged that the algorithms \cite{Shor1994} expected to provide the most significant advantage over their classical counterparts will require error correction, where information is encoded in logical qubits and processors need millions of physical qubits to operate \cite{Fowler2012, Lekitsch2017}. Scalability is thus an unavoidable objective for any quantum computation technology platform.

Electron spin qubits in gate-defined silicon quantum dot (QD) devices are a leading  platform for realizing large-scale quantum computers. Silicon QDs exhibit relatively long coherence times \cite{Veldhorst2014}, are able to operate at temperatures above 1~K \cite{Yang2020, Petit2020}, and can utilize traditional very large-scale integration (VLSI) fabrication processes \cite{Veldhorst2017,Hutin2018}. 
The feasibility of universal quantum computing in silicon has been established through the demonstration of high-fidelity single \cite{Veldhorst2014,Yoneda2018,Yang2019} and two \cite{Veldhorst2015,Huang2019,Xue2021} qubit gates, with the focus now on scaling up to NISQ devices and beyond to fault-tolerant systems \cite{Veldhorst2017}. 

A critical requirement along this path is the ability to deliver microwave signals, which are needed to control spin qubits, across the entire quantum chip. Current methods for controlling electron spin qubits in devices include direct magnetic driving using on-chip transmission lines (TL) \cite{Dehollain2013} and electrically driven spin resonance (EDSR) \cite{Pioro2008, Kawakami2014, Takeda2016, Watson2018, Yang2020}, both of which have proven useful in small scale (1 to 10 qubit) device demonstrations. However, as the number of qubits scales up, heating and design complexity issues will need to be resolved \cite{Li2018,Struck2020} for these approaches to be practical. 

Global control \cite{Kane1998} was an early technique proposed for spin qubits, offering a scalable solution to the problem of delivering microwave signals. Here, a global and uniform magnetic field is applied across the entire quantum chip in order to drive each individual qubit \cite{Veldhorst2017, Hill2015, Vahapoglu2021}. Qubit manipulation is activated by locally applied electric fields to shift their individual resonance frequencies into and out of resonance with the global field \cite{Laucht2015}. This technique does not suffer the same scalability concerns as the TL or EDSR approaches, since no high frequency lines or microwave currents directly pass through the chip. Furthermore, it requires only a single microwave signal generator.

Despite the appeal of global control, owing to its simplicity, single spin resonance using a global field was only achieved recently. A 3D microwave dielectric resonator (DR) made from potassium tantalate (KTaO$_3$ or KTO) \cite{Blank2003,Vahapoglu2021,Vallabhapurapu2021} was used to create a global, off-chip magnetic field and perform electron spin resonance (ESR) of single spins in a planar SiMOS device \cite{Vahapoglu2021}. The microwave magnetic field generated by the DR produced incoherent mixing of spin states in a double quantum dot (DQD), which was detected with a single electron transistor through a process known as spin to charge conversion \cite{Petta2005}. However, the coherent control of spin qubits using a global field has remained an outstanding challenge that must to be addressed to fully establish the feasibility of applying this technique in large-scale quantum computers.

Here we present the coherent control of single electron spin qubits in a planar SiMOS DQD device using a global magnetic microwave field generated off-chip by a KTO dielectric resonator. We measure Rabi frequencies and report on the coherence properties of both spin qubits. We compare the noise spectrum seen by the qubits in this DR-driven device to that observed by traditional TL-driven qubits and conclude that this new scalable approach does not diminish the performance of the qubits. This work demonstrates the potential of global control using dielectric resonators as a scalable qubit control technique.

\section*{Single Spin Resonance}\label{section::ESR}

	\begin{figure*}[!t]
	  \centering
	  \includegraphics{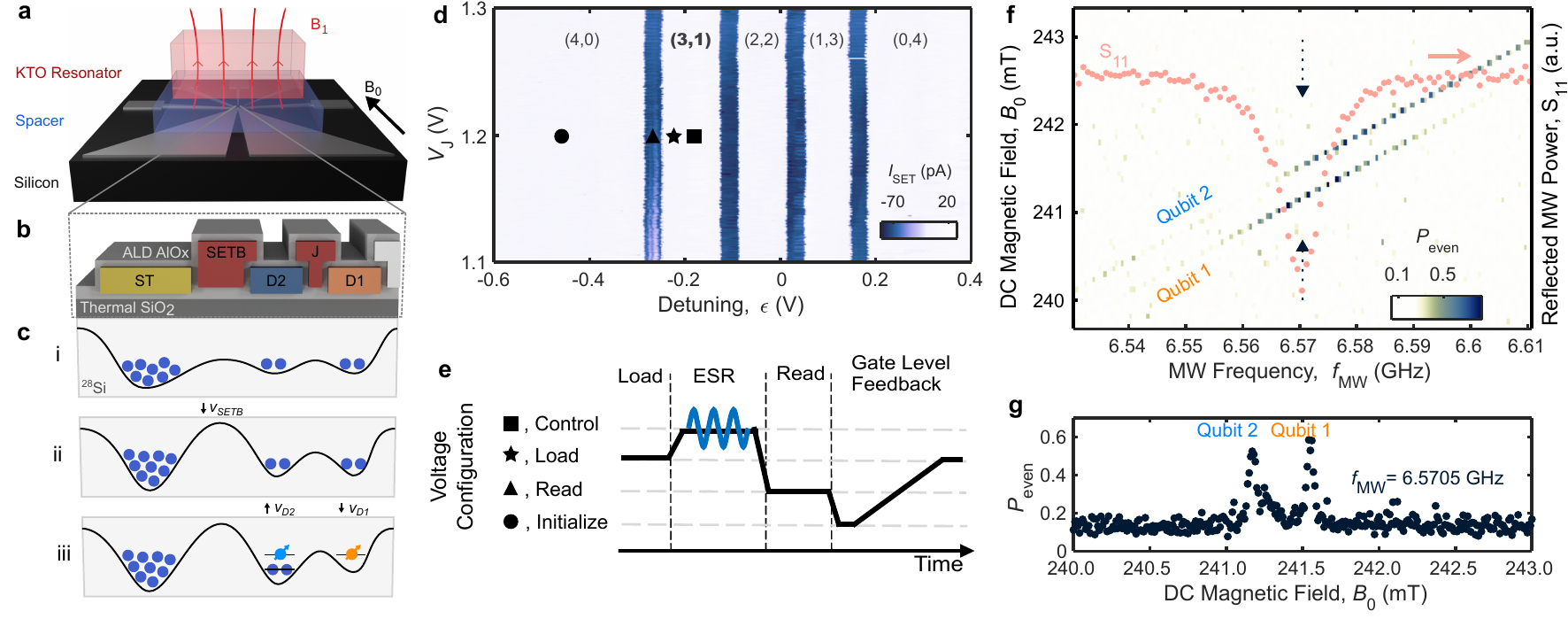}
	  \caption{\textbf{Device stack and electron spin resonance (ESR).}
		\textbf{a}, A 3D render of the global control device stack used in our experiments, including the silicon quantum nanoelectronic device (bottom, black), sapphire dielectric spacer (middle, blue) and potassium tantalate (KTO) dielectric microwave resonator (top, pink).
		\textbf{b}, A schematic cross-section through the measured silicon QD device showing the 3D structure of the gates, enriched silicon-28 substrate and insulating oxide layers.
		\textbf{c}, Steps for preparing the isolated mode measurements, depicted with conduction band profiles at the interface of the \textsuperscript{28}Si substrate. The preparation consists of three phases: loading (i), isolation (ii), and initialization (iii) (see text for more details).
		\textbf{d}, Charge stability diagram measured in isolated mode. Four charge transitions occur while the detuning voltage $\epsilon$ is swept from -0.6~V to 0.4~V, indicating that there are 4 electrons in the double dot system. Black geometric symbols (explained in panel e) show the voltage configurations used in the ESR experiments.
		\textbf{e}, Pulsing scheme for the ESR measurements. The DQD is initialized as a singlet state $(\ket{\downarrow\uparrow}+\ket{\uparrow\downarrow})/\sqrt{2}$ in the (3,1) charge configuration point near the (3,1)-(4,0) transition. It is then pulsed deeper into the (3,1) region (square) and a microwave pulse is applied to the dielectric resonator, generating an alternating magnetic field, $B_1$, which can rotate the spins if they are in resonance with the field. Readout is performed in the Pauli Spin Blockade region (triangle) and reveals if the system is in an even or odd state (see the text for more detail). Finally, a diabatic ramp from (4,0) (circle) to (3,1) (star) is applied in order to both re-initialize the DQD in a singlet state (see Supplementary Information) and implement gate-level feedback.  
		\textbf{f}, Even state probability as a function of $f_{\rm MW}$ and $B_0$, revealing two ESR peaks that shift with magnetic field, consistent with two spin qubits occupying the double dot system. The peaks are labeled Qubit 1 and Qubit 2. An $S_{11}$ reflection measurement (pink circles) probed via the coaxial loop coupler is superimposed over the 2D map. The ESR pulse duration is fixed at $1.5~\mu$s.
		\textbf{g}, Even state probability as a function of $B_0$ when $f_{\rm MW}$ corresponds to the center frequency of the DR resonance (dotted lines with arrows in panel f). The pulse duration is sufficiently long to make the ESR drive incoherent ($ 25~\mu $s), causing the spin states to become completely mixed and resulting in a peak amplitude of $P_{\rm even} = 0.5$ for both resonances.
		}
	  \label{fig:1}
	\end{figure*}

The device studied here employs a nominally identical stack as reported previously \citealp{Vahapoglu2021} (see Fig.~\ref{fig:1}a), which consists of a 0.7~mm~$\times$~0.55~mm~$\times$~0.3~mm rectangular KTO prism positioned above a silicon quantum nanoelectronic device. A 0.2~mm thick low-loss dielectric sapphire spacer is positioned between the DR and device to isolate them from one another. The fundamental mode of the DR produces an alternating magnetic field ($B_{1}$) out of the qubit plane and in a direction perpendicular to the DC magnetic field ($B_{0}$), as shown in Fig.~\ref{fig:1}a. We use the $B_{1}$ field to control the spin state of the qubits via magnetic resonance. A coaxial loop coupler, through which the microwave power is inductively coupled to the DR, is placed above the stack\cite{Vahapoglu2021} (not shown).
 
The qubit device is a metal-oxide-semiconductor (MOS) DQD formed in an isotopically enriched silicon-28 substrate (50 ppm residual \textsuperscript{29}Si), whose cross-sectional view is depicted in Fig.~\ref{fig:1}b. This is in contrast to previous work \cite{Vahapoglu2021} which employed a natural silicon substrate with a $\sim~4.7\%$ abundance of \textsuperscript{29}Si nuclei, which produced strong dephasing of the electron spins. A scanning electron microscope (SEM) image of an identical device to the one measured is also provided in the Supplementary Information. The QDs are electrostatically defined by a palladium (Pd) multi-layer gate stack architecture in which different layers are electrically isolated by atomic-layer-deposited (ALD) aluminium oxide (AlO\textsubscript{x}), since Pd does not form its own native oxide \cite{Brauns2018}. A thermally grown SiO\textsubscript{2} layer above the silicon substrate prevents any current leakage between the gate electrodes and the substrate. The device consists of a single electron transistor (SET), which is used as a charge sensor, with a top gate (ST) for tuning its charge accumulation and sensitivity, two plunger gates (D1-D2) for forming the quantum dots and setting their charge occupations, two barrier gates that control the coupling between the dots (J) or between dot 2 and the SET island (SETB), and confinement gates (CB1-CB2) to laterally confine the dots (not shown in Fig.~\ref{fig:1}b, see the SEM image in the Supplementary Information).

Measurements in this work have been carried out in a configuration referred to here as \textit{isolated mode}, where the electrons inside the double dot system are electrically isolated from the nearby electron reservoirs, as employed elsewhere\cite{Yang2020} (see Supplementary Information). The steps required to prepare this configuration are depicted in Fig.~\ref{fig:1}c and entail three main stages. In the loading phase (i), electrons are introduced to the double dot system via the SET, which is also coupled to an electron reservoir (not shown). The number of electrons loaded can be tuned using the D1, D2, and J gates. The system is then isolated (ii) by raising the potential underneath the gate SETB. Finally, the desired charge occupation is initialised (iii) by setting the plunger gate (D1 and D2) voltages appropriately.

Figure~\ref{fig:1}d shows a charge stability diagram measured in isolated mode with a double lock-in technique \cite{Yang2011}. The four vertical blue lines indicate that 4 electrons are trapped inside the DQD system, tunneling between the dots depending on the value of the voltage detuning ($\epsilon = V_\text{D1}-V_\text{D2} $). A more positive $\epsilon$ favours electron occupation under gate D1, whilst a more negative $\epsilon$ favours occupation under D2. The absence of additional lines on both sides of the diagram confirms that only 4 electrons have been loaded in the DQD system. The charge configurations are labelled as (N2, N1) where N1(2) refers to the number of electrons under D1(2). 

In the following spin measurements we focus on the (3,1) charge configuration (emphasised in Fig.~\ref{fig:1}d), which provides an equivalent spin state to (1,1) since the first two electrons under D2 form a spin-zero closed shell and do not interact with the remaining electrons in the system (see panel iii in Fig.~\ref{fig:1}c). Fig.~\ref{fig:1}e depicts the pulse scheme applied during the measurements. The system is first initialised in a spin singlet state $(\ket{\downarrow\uparrow}+\ket{\uparrow\downarrow})/\sqrt{2}$ (see Supplementary Information). Following this, the system is plunged into the middle of the (3,1) region whilst a microwave signal is applied to the loop coupler for a period of time in order to flip the spins in the DQD with the global $B_1$ field. If the frequency of the $B_{1}$ field matches one of the qubit frequencies, i.e. $f_\text{MW} = g\mu_\text{B}B_0/h$ (where $g$ is the electron g-factor, $\mu_\text{B}$ the Bohr magneton and $h$ Planck's constant), the resonant spin qubit will flip between the $\ket{\uparrow}$ and $\ket{\downarrow}$ states. The resulting DQD spin state is then translated into a charge state in the Pauli Spin Blockade (PSB) region, measured using the SET, as either a singlet or triplet state\cite{Petta2005}. However, we have confirmed that the blockade is restricted to even states instead of all triplet states (in what is known as parity readout~\cite{Seedhouse2021pauli}). Therefore, the readout effectively unveils an even or odd state.

In Fig.~\ref{fig:1}f we plot the measured DQD even state probability ($P_\text{even}$) after applying the measurement sequence (Fig.~\ref{fig:1}e), as we scan both the $B_0$ field and the microwave frequency $f_\text{MW}$. Electron spin resonance signatures are detected for the spins in both dots (diagonal features in Fig.~\ref{fig:1}f), where it is observed that the qubit frequencies shift linearly as a function of $B_{0}$, confirming the features are indeed electron spin resonance (ESR) peaks. By monitoring how the qubit frequencies change due to Stark shifts from voltages applied to the different gates (see analysis in the Supplementary Information), we deduce that the spin resonance signal at the higher frequency (for a fixed $B_0$) corresponds to the spin under gate D1, whilst the other resonance belongs to the spin under D2. We thus label the resonances as `Qubit 1' and `Qubit 2', respectively. We perform an $S_{11}$ reflection measurement on the dielectric resonator via the loop coupler and superimpose it on Fig.~\ref{fig:1}f (pink circles). The amplitude of the ESR peaks are correlated with the $S_{11}$ DR resonance, indicating that the $B_{1}$ field generated by the KTO resonator indeed drives the qubits \cite{Vahapoglu2021}. In Fig.~\ref{fig:1}g we show a one-dimensional slice of the qubit resonances taken at the DR center frequency, measured along the black dotted arrows marked in Fig.~\ref{fig:1}f.

These results are consistent with recent observations in a natural silicon device \cite{Vahapoglu2021}, confirming that this new spin control technology is device independent. Next, we exploit the long spin coherence times available in enriched \textsuperscript{28}Si to enable \textit{coherent} qubit control using the off-chip $B_1$ field. 

\section*{Coherent Qubit Control}\label{section::T2}

\begin{figure*}[ht!]
	\centering
	\includegraphics[]{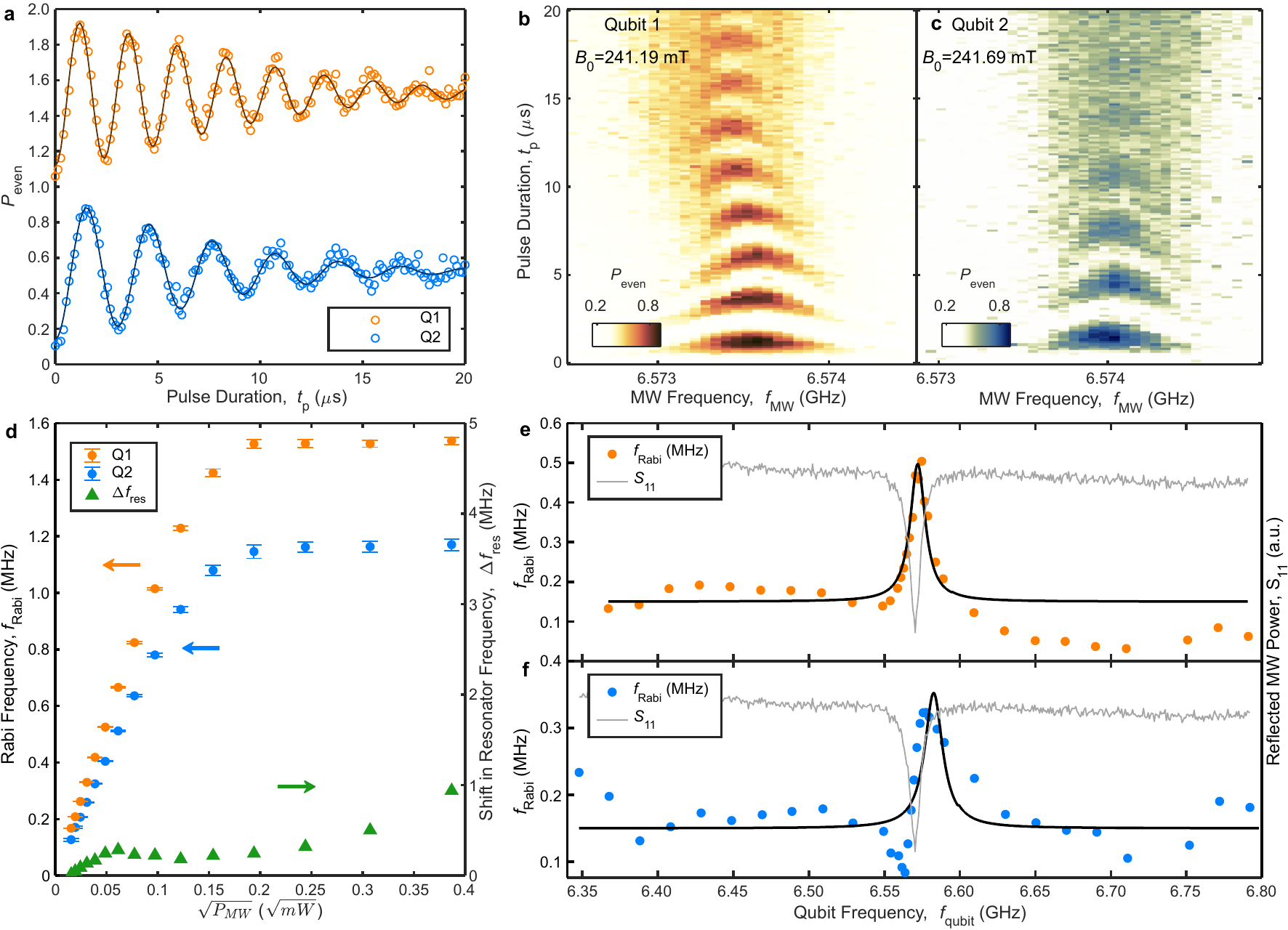}
	\caption{\textbf{Coherent control.}
		\textbf{a}, Coherent Rabi oscillations for both qubits. The Qubit 1 measurement is offset by $1.0$ for clarity.
		\textbf{b-c}, Rabi chevrons for Qubit 1 (b) and Qubit 2 (c). The DC magnetic field is tuned to shift the qubits close to the DR center frequency.
		\textbf{d}, Rabi frequency vs. MW power applied to the coaxial loop coupler. For low powers, the Rabi frequencies of both qubits are linearly proportional to the square root of the power, as expected (see text for more details). The Rabi frequencies begin to saturate at an input power of $\sim~20~\mu$W, as discussed in the text. We plot the shift in the DR frequency from its low-power value (green triangles) as a function of the applied microwave power, measured in a continuous wave experiment.
		\textbf{e-f}, Rabi frequencies vs. qubit frequencies for Qubit 1 (e) and Qubit 2 (f). For each data point (solid circles), the DC magnetic field is tuned appropriately in order to shift the qubits to the desired frequency, then $f_\text{Rabi}$ at that qubit frequency is measured. The error bars are not shown since they all lie within the extent of the data point markers. The DR $S_{11}$ measurement from Fig.~\ref{fig:1}f is superimposed (grey line) on these panels for ease of comparison. The region of enhanced Rabi frequencies overlaps with the DR response, confirming that magnetic resonance via the KTO DR is the primary mechanism for driving spin rotations in these regions. Black solid lines are Lorentzian fits to the $ f_\text{Rabi} $ distributions.
	}
	\label{fig:2}
	
\end{figure*}

We measure the even state probability $P_{\rm even}$ as a function of the applied ESR pulse length ($t_\text{p}$), with $f_{\rm MW}$ and $B_{0}$ chosen to satisfy the resonance condition $f_\text{MW} = g\mu_\text{B}B_0/h$ for each qubit. The result is plotted in Fig.~\ref{fig:2}a and clearly demonstrates coherent Rabi oscillations. Repeating this measurement as a function of $f_{\rm MW}$ for a fixed $B_0$ (see Figs.~\ref{fig:2}b-c) we observe Rabi chevron patterns for both qubits, where faster driving occurs as $f_{\rm MW}$ becomes detuned from resonance with the qubits and is accompanied by a reduction in the oscillation visibility. 

The relation between the $B_{1}$ field generated by the DR and the applied microwave power is given as $B_{1} = C\sqrt{P_\text{MW}}$, where $C$ is the conversion factor and $P_\text{MW}$ is the power \cite{Vahapoglu2021,Vallabhapurapu2021}. Therefore, we expect $f_{\rm Rabi} = g\mu_\text{B}B_1/h$ to have a linear dependence on $\sqrt{P_\text{MW}}$. This is investigated in Fig.~\ref{fig:2}d, where for low powers $f_\text{Rabi}$ indeed increases linearly with respect to $\sqrt{P_{\rm MW}}$ with an average conversion factor extracted from the slopes of this plot of $C \approx 11~\mu\text{T}/\sqrt{\mu\text{W}}$. 
For powers exceeding $P_\text{MW} \approx 20~\mu$W  the Rabi frequency begins to saturate for both qubits. We probe the shift in the resonance frequency of the KTO resonator (relative to its low-power value) as a function of $P_\text{MW}$ and find that the change is minimal ($<~1$~MHz) over the range of powers measured. We therefore conclude that the observed saturation in $f_{\rm Rabi}$ is not caused by power-induced shifts of the DR frequency, for example due to heating of the KTO dielectric \cite{Vallabhapurapu2021}. The saturation could be indicative of another driving mechanism (e.g. via electric fields \cite{Corna2018,Huang2017}) that becomes significant at higher powers and competes with the $B_1$ control field from the DR. Further work is warranted to understand the mechanism behind the Rabi frequency saturation.

Next we explore how $f_\text{Rabi}$ varies with respect to the qubit resonance frequencies ($f_\text{qubit}$) in order to quantify the effect of the resonator more accurately. We have measured Rabi oscillations for different $f_\text{qubit}$ values (setting $B_{0} = hf_\text{qubit}/g\mu_\text{B}$) around the fundamental mode of the DR, and plot $f_\text{Rabi}$ against $f_\text{qubit}$ for both qubits in Figs.~\ref{fig:2}e and f. It is clear from these measurements that $f_\text{Rabi}$ is enhanced for qubit frequencies corresponding with the DR resonance (as can be seen from the superimposed $S_{11}$ measurement), which implies that the qubits are primarily driven by the KTO resonator in these regions. Compared to the off-resonant drive, $f_\text{Rabi}$ is enhanced at the center of the DR resonance by a factor of 3.3 for Qubit 1 and 2 for Qubit 2. We note that a factor 3.3 enhancement in $B_1$ corresponds to a $> 10$ times lower power requirement for a given field strength. There is also a residual drive that is present across all qubit frequencies, which we believe could originate from several sources. Microwaves may couple into a broadband transmission line (unused in this experiment, see Supplementary Information) that is terminated 200~nm from the DQD, producing a weak $B_1$ field across a large bandwidth. In addition, microwave currents that are induced in the metal gate electrodes may produce magnetic or electric fields that result in weak ESR or EDSR \cite{Huang2017}.

\section*{Coherence Time Measurements}\label{section::T2}
	\begin{figure}[t!]
	\centering
	\includegraphics[]{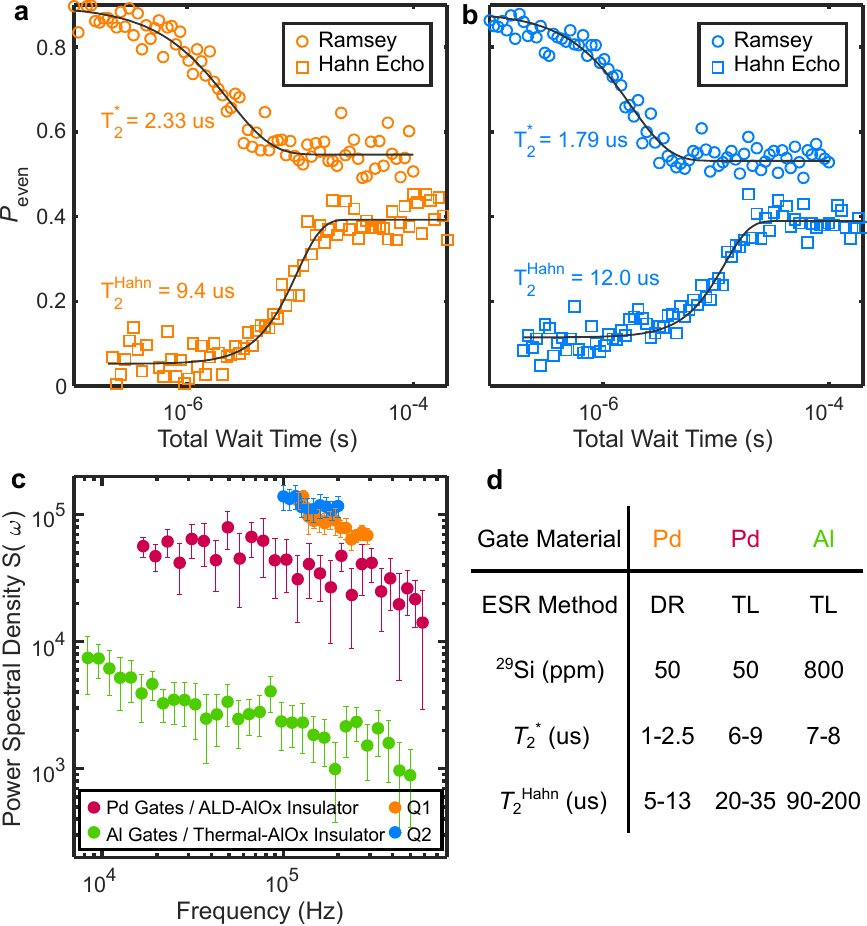}
	\caption{\textbf{Coherence time measurements}.
		\textbf{a-b}, Spin coherence times measured via Ramsey free induction decay ($ T^{*}_{2} $) and Hahn echo ($ T^{Hahn}_{2} $) experiments for Qubit 1 (a) and Qubit 2 (b). The Hahn echo data are offset by -0.1 for clarity.
		\textbf{c}, CPMG noise spectroscopy measurements taken from three devices having a similar gate layout (Orange: Our device - Qubit 1, Blue: Our device, Qubit 2, Purple: Device A - Pd gate electrodes with an ALD-AlO\textsubscript{x} insulator, Green: Device B - Al gate electrodes with a thermal-AlO\textsubscript{x} insulator). ${\rm S}(\omega)$ is the power spectral density of the qubit frequency noise.
		\textbf{d}, Table comparing the coherence times of the devices examined in panel c.
	}
	\label{fig:3}
	\end{figure}

We investigate the coherence times of the qubits by performing Ramsey free induction decay and Hahn echo experiments. The data measured in these experiments are shown in Figs.~\ref{fig:3}a and b for Qubit 1 and 2, respectively. To extract the $T_{2}^{\rm *}$ times, we fit the Ramsey data to exponential decay functions of the form $P_{\rm even} = A e^{-(t/T_{2}^{\rm *})^n} + B$, where the parameters $A$ and $B$ are related to the measurement visibility and $n$ is the decay exponent, typically ranging between 1 and 2 \cite{Paladino2014}. The fits reveal $T_{2}^{\rm *} = 2.33\pm0.35~\mu$s for Qubit 1 and $T_{2}^{\rm *} = 1.79\pm0.20~\mu$s for Qubit 2. Similarly, the Hahn echo measurement results are fit with the function $P_{\rm even} = A (1-e^{-(t/T_{2}^{\rm Hahn})^2}) + B$, yielding $T_{2}^{\rm Hahn} = 9.4\pm1.0~\mu$s for Qubit 1 and and $T_{2}^{\rm Hahn} = 12.0\pm1.2~\mu$s for Qubit 2.

The reported spin coherence times are 1-2 orders of magnitude smaller than the best measured values in silicon MOS devices \cite{Veldhorst2014}. In order to determine if this is related to the DR, we make a comparison of the coherence and noise properties for typical silicon qubit devices constructed with palladium gates and aluminium gates (Figs.~\ref{fig:3}c and d). We first compare Carr-Purcell-Meiboom-Gill (CPMG) noise spectroscopy measurements \cite{Alvarez2011} from our device and two other devices (one containing palladium gates and the other aluminium gates) having nearly identical gate layouts, but where the qubit control signals are delivered by conventional on-chip TLs (Fig.~\ref{fig:3}c). The TL device with Al gate electrodes (green circles) has a thermally grown AlO\textsubscript{x} gate insulator, while the other TL device (purple circles) is made from exactly the same materials as the current device (Pd gates with an ALD AlO\textsubscript{x} insulator). The results show that the devices with Pd gate electrodes and ALD insulators have similar noise floors, which is an order of magnitude higher than that of the Al gate device, despite the residual \textsuperscript{29}Si concentration (50\,ppm) being considerably lower in the Pd devices than the one containing Al gates (800\,ppm).

In Fig.~\ref{fig:3}d we compare coherence times of the TL devices and the DR device measured here. The times for the Pd gate device with an on-chip TL (dark red) are comparable (i.e. within a factor of $2-3$) to those reported for the current device with the off-chip DR (orange), while the Al gate device (green) has an order of magnitude higher $T_{2}^{\rm Hahn}$, which is consistent with the CPMG noise spectroscopy results in Fig.~\ref{fig:3}c. We believe that the higher decoherence of the qubits in the Pd devices is most likely related to the materials used, where charge noise arising from the ALD-grown AlO\textsubscript{x} gate oxide layers \cite{Connors2019} is a potential source. The factor $2-3$ difference in coherence times for the Pd devices (with and without DR) could be due to device variability, or perhaps because the oxide charge is disturbed over a larger area in the DR experiment. Variability in $T_{2}^{\rm Hahn}$ will be investigated in the future by measuring additional devices. Ultimately, we believe that moving to an Al device with thermally-grown AlO\textsubscript{x} should substantially improve the qubit coherence times.

\section*{Discussion}
We have demonstrated, for the first time, the coherent control of spin qubits in a nanoelectronic device using a globally-applied magnetic field, achieving Rabi frequencies greater than 1~MHz for an input power of approximately $20~\mu\text{W}$. We also report a comparative analysis of the coherence and noise properties of devices where ESR is driven via on-chip transmission lines (local control) to those employing an off-chip dielectric resonator (global control). The coherence times of the local and global control devices made from identical materials (Pd gates with an ALD oxide) are within a factor of 3, but differ by an order of magnitude from that measured in a local control device made with an Al gate stack, leading us to suspect the device materials as the origin of noise. In future work, we plan to apply this off-chip DR control technique to a device made with Al gate electrodes and thermally grown AlO\textsubscript{x} gate insulators, which we expect will result in longer spin coherence times.

Another avenue for improvement is the DR quality factor ($Q_\text{i} = 780$), which is currently limited by losses in the device. The material-limited quality factor for KTO DRs is approximately two orders of magnitude larger ($Q_\text{i} > 60,000$) \cite{Vahapoglu2021} -- reaching this limit would mean the power could be reduced a hundredfold for a given $B_1$ amplitude, with corresponding lower levels of undesired disturbance, for example due to unintended microwave loops or resonances in the device. 

The improvements in global control hardware developed here should be accompanied by efforts to design and optimize pulse protocols for implementing high fidelity single qubit gate operations \cite{Hansen2021,Hansen2021global}. In addition, qubit operations such as initialisation, readout and two-qubit entangling gates must be harmonised with the presence of the continuously driven microwave field generated by a high-Q resonator. Recent work in this direction \cite{Seedhouse2021} shows that high fidelity gates should be possible with realistic experimental parameters.

Our work shows that delivering microwave signals to spin qubits in a quantum processor -- something that has so far been seen as a major challenge and drawback of the platform -- can be elegantly resolved by means of global control using a KTO dielectric resonator. Demonstrating off-chip coherent control of spin qubits brings the prospect of large-scale spin-based quantum computers one important step closer.

\bibliographystyle{naturemag}
\bibliography{OffChipReferences}

\section*{Acknowledgments}
\paragraph*{Funding:} The authors acknowledge support from the Australian Research Council (DE190101397, FL190100167 and CE170100012), the US Army Research Office (W911NF-17-1-0198) and the NSW Node of the Australian National Fabrication Facility. The views and conclusions contained in this document are those of the authors and should not be interpreted as representing the official policies, either expressed or implied, of the Army Research Office or the US Government. E.V. and J.P.S.-S. acknowledge support from Sydney Quantum Academy. The authors thank Peter Becker for the preparation of the isotopically-purified silicon substrate. 
\paragraph*{Author contributions:} E.V. and J.P.S.-S. performed the experiments. J.P.S.-S. and J.J.P. designed and fabricated the DR. T.D. designed the sample enclosure and assisted with the DR characterization. W.H.L. and F.E.H. fabricated the silicon device. M.L.W.T., N.V.A. and H.-J.P. prepared and characterized the isotopically-purified silicon substrate. E.V., J.P.S.-S., R.C.C.L, J.D.C. and A.S. analysed the data. C.H.Y., T.T., A.L. and A.S. contributed to discussions on the experimental results. E.V., J.P.S.-S., J.J.P. and A.S.D. wrote the manuscript with input from all authors. J.J.P. and A.S.D. supervised the project.
\paragraph*{Competing Interests:} J.J.P. and A.S.D. are inventors on a patent related to this work (PCT AU2020/051239) filed by the University of New South Wales with a priority date of 15 November 2019. All other authors declare that they have no competing interests.

\end{document}